\documentclass[12pt]{article}
\usepackage{graphicx}
\usepackage{rotating}
\usepackage{multirow}
\usepackage{amssymb}
\usepackage{epstopdf}
\usepackage[usenames,dvipsnames]{color}

\newcommand{\xxi}{$21^{\rm st}$}

\def\deg{$^{\circ}$}

\title{Delayed Seasonal Cycle and African Monsoon in a Warmer Climate}

\author
{Michela Biasutti,$^{1\ast}$  Adam H. Sobel$^{2}$\\
\\
\normalsize{$^{1}$Lamont-Doherty Earth Observatory of Columbia University,}\\
\normalsize{61 Route 9W, Palisades, NY 10964,  USA}\\
\normalsize{$^{2}$Department of Applied Physics and Mathematics}\\
\normalsize{and Department of Earth and Environmental Sciences,  Columbia University,}\\
\normalsize{New York, NY 10027,  USA}\\
\\
\normalsize{$^\ast$To whom correspondence should be addressed; E-mail:  biasutti@ldeo.columbia.edu.}
}

\date{}

\begin{document} 


\baselineskip18pt


\maketitle


{\bf 
Increasing greenhouse gases will change many aspects of the Earth's climate, from its annual mean to the frequency of extremes such as heat waves and droughts\cite{ipcc4}.  Here we report that the current generation of climate models predicts a delay in the seasonal cycle of global rainfall and ocean temperature in response to increasing greenhouse gases, with important implications for the regional monsoons.  In particular, the rainy season of the semi-arid African Sahel is projected to start later and become shorter: an undesirable change for local rainfed agriculture and pastoralism. Previous work has highlighted the uncertainty in this region's response to anthropogenic global warming: summer rainfall is predicted either to decrease or increase by up to 30\% depending which model is used\cite{hel+al05, bia+al08, coo08}. The robust agreement across models on the seasonal distribution of rainfall changes signifies that the onset date and length of the rainy season should be more predictable than annual mean anomalies.  
}

Changes in the timing of the North African monsoon are a manifestation of a delay of the global seasonal cycle in response to greenhouse gases (GHG), thus we first describe that global response. We use the Coupled Models Inter-comparison Project (phase 3, CMIP3) dataset, which provided the basis for the latest report of the Intergovernmental Panel on Climate Change (IPCC).  Coherent spatial patterns and temporal evolution of the seasonal cycle of global sea surface temperature (SST) and precipitation are captured by empirical orthogonal functions (EOFs) and principal components (PCs) of the monthly climatology, respectively\cite{kut67}.
\begin{figure}
\begin{center}
\includegraphics[scale=.8]{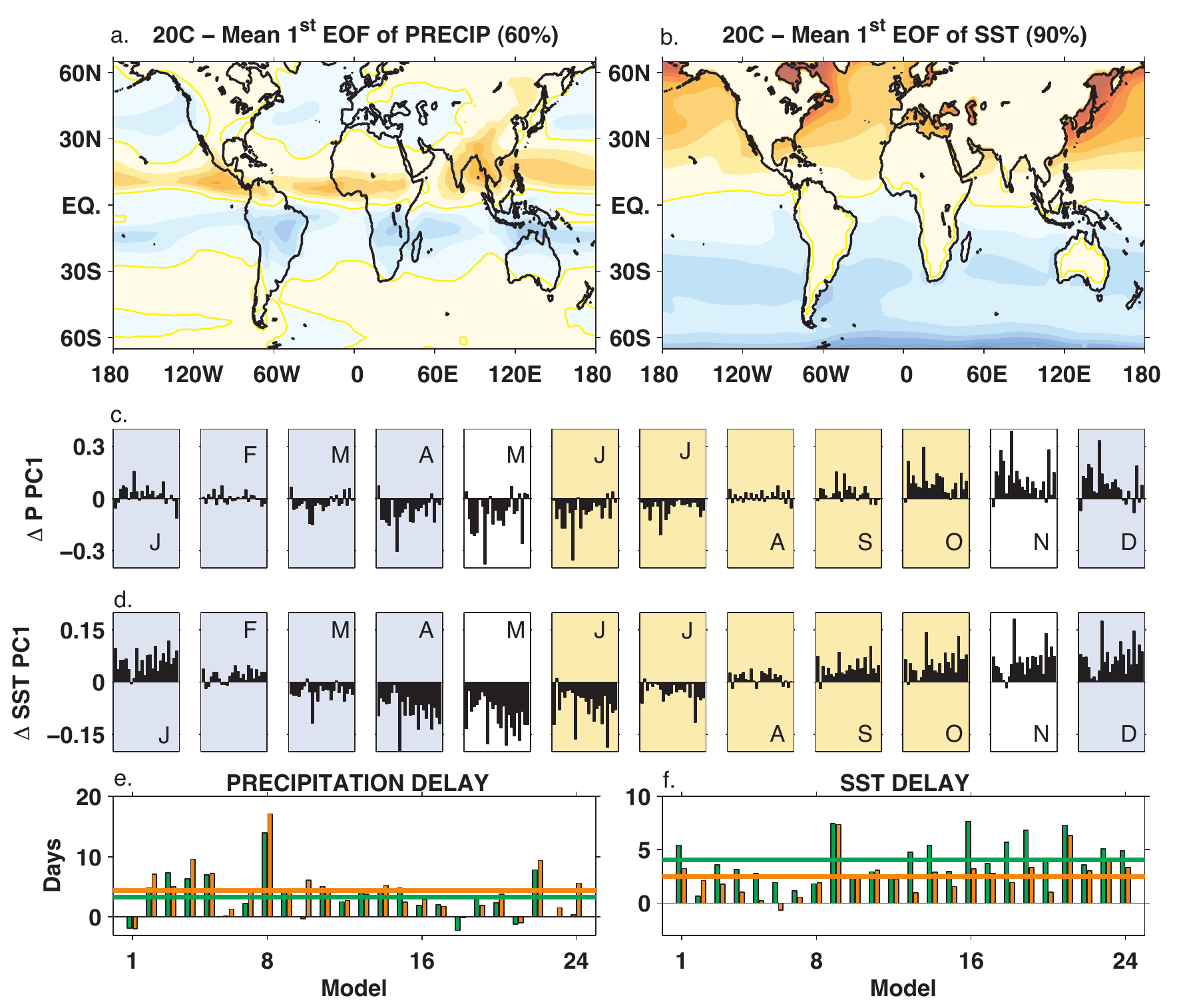}
\caption{{\footnotesize {\bf Annual Cycle of Precipitation and Sea Surface Temperature in the 20th and 21st Century.} Top: First CMIP3 ensemble mean EOF of the 20th century climatology of (a.) precipitation and (b.) SST. Middle panels: 21C-20C difference in PC1 of (c.) precipitation and (d.) SST in each calender month and for each of the 24 CMIP3 models. Blue and orange background shadings indicate months in which PC1 is negative or positive, while no shading indicates a transition month. Bottom: 21C-20C delay in the timing of zero-crossing of PC1 of (e.) precipitation and (f.) SST; green bars refer to the spring crossing, orange bar to the autumn crossing.}}
\label{EOFs}
\end{center}
\end{figure}
Precipitation variations with an annual period are described by two EOF/PC pairs (explaining 60\% and 20\% of the variance in the climatology).   EOF1 of precipitation (Figure~1a) captures the observed shift of rainfall to the summer hemisphere in the monsoon regions and in the oceanic Inter-Tropical Convergence Zone (ITCZ). EOF2  (Figure S1) describes a north-south dipole in the marine ITCZs, like EOF1, but emphasizes the contrast between continental and oceanic regions. The annual cycle of SST is fully captured by EOF1 (explaining 90\% of variance, Figure 1b). It is largest at high latitudes, especially east of Asia and North America, where heat is lost to continental air in winter, and in regions of seasonal sea ice and smaller in the tropics, where small SST changes can nonetheless effectively drive changes in local circulation and precipitation\cite{chi+al01}.  The first PCs (not shown) of precipitation and temperature show a peak in August, when the northern hemisphere (NH) ocean is the warmest and precipitation reaches farthest to the North. The phase offset between SST and NH insolation is due to the high heat capacity of the ocean mixed layer; that  PC1 of rainfall is equally offset underscores how SST controls tropical precipitation. 

The differences in PC1s between the last twenty-five years of the 21st century and of the 20th century (the 21C-20C anomalies) are in quadrature with the PC1s themselves, indicating a shift in the evolution of the seasonal cycle of both SST and rainfall during the \xxi\ century (Figure~1c,d); this behavior is robust across the CMIP3 ensemble. By fitting PC1 to a sinusoid, we quantify the shift as 3.2 days for SST and 3.7 days for precipitation. Alternatively, the times when PC1 crosses the zero line in spring and fall provide an estimate of the delay and the change in the length of the seasons. On average, PC1 of SST crosses the zero line with a 4.1-day delay in spring and a 2.5-day delay in fall; for precipitation the estimates are 3.3 and 3.7 days, respectively (rainfall PC2 indicates a shift of the same order,  Figure~S1). Variations across the CMIP3 ensemble in the estimates of the delay are especially small in the case of SST (Figure~1f), suggesting that the difference between the delay in spring and fall might be robust and that, by this measure, the NH summer would become shorter by a day and a half. Of course, given the expected annual mean increase in temperature, the warm season as defined by a fixed threshold will actually be longer than in the past.
%
%
The global nature of the delay is not an artifact of EOF analysis, but appears in the evolution of the 21C-20C anomalies of both rainfall and SST (Supplementary Material and Figure S2).


\begin{figure}
\begin{center}
\includegraphics[]{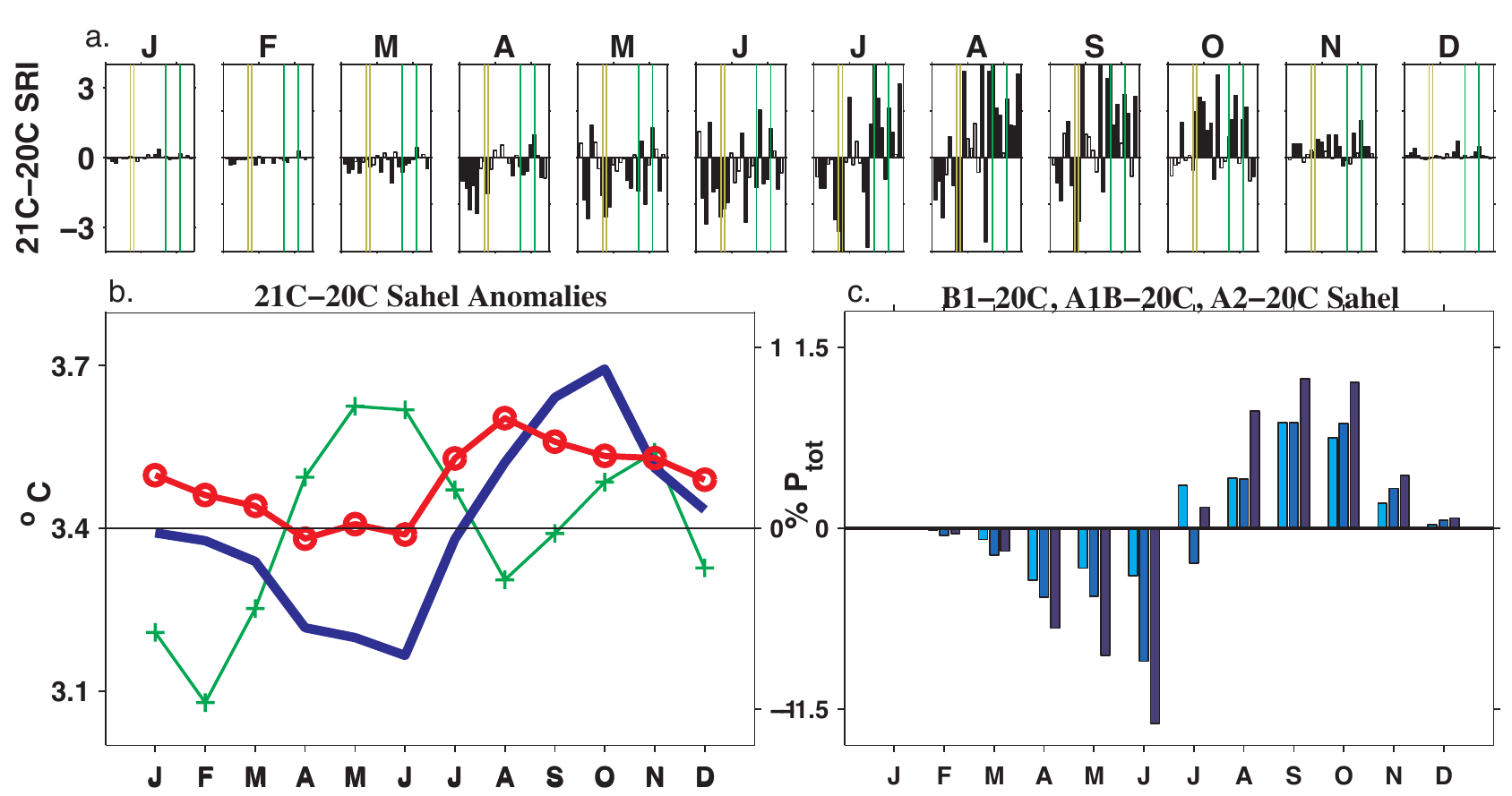}
\caption{{\footnotesize  {\bf 21C-20C Anomalies in Sahel Rainfall.} (a.) Monthly (January through December) anomalies for individual CMIP3 models. The thin brown (green) lines indicate the GFDL models (MIROC\_medres and NCAR CCSM3). (b.) Ensemble Mean change in Sahel precipitation (solid blue) and evaporation (red circles), expressed as a percentage of the 20C total annual precipitation (right axis) and change in surface temperature (green pluses, left axis). (c.) Precipitation anomalies as a function of month in three scenario with stronger concentrations of GHG: B1-20C (cyan), A1B-20C (light blue) and A2-20C (dark blue). Anomalies are in percentage of 20C total annual precipitation.}}
\label{Fig2}
\end{center}
\end{figure}

In the Sahel, the delay appears robust in spite of high uncertainty in the projections of annual mean rainfall anomalies\cite{hel+al05, bia+al08, coo08}.  Rainfall anomalies are predominantly negative across the CMIP3 models at the beginning of the rainy season (in May and June), but become positive at the end of the season, in October (Figure 2).  Because there is great variability in when models switch from negative to positive anomalies, the overall change in annual mean rainfall is model dependent, though anomalies at the beginning and end of the rainy season are robust.  In fact, differences in the delay are largest among the models previously singled out for their outlying projections of Sahel rainfall\cite{hel+al05, bia+al09}. The GFDL models, which project extreme drying, simulate negative anomalies up to September, while the medium-resolution MIROC model (with the wettest projection) does so only up to June; the NCAR model CCSM3 is the only model in the CMIP3 dataset that simulates positive rainfall anomalies throughout the year.  

The spring reduction in precipitation is not matched by a reduction in evaporation (Figure~2b): anomalous evaporation is always positive and precipitation minus evaporation (P-E) anomalies are negative for most of the year (except fall), potentially leading to drier soil. Positive evaporation anomalies independent of rainfall anomalies have been associated with anthropogenic warmer temperatures in other semi-arid regions\cite{nic04}.  
In spring, the cloudiness reduction associated with negative rainfall anomalies is not offset by higher evaporation, so that surface temperatures are higher than would have been otherwise.  Temperatures in the Sahel are projected to be an average of 3.4\deg C warmer, with possibly dire consequences for agriculture\cite{bat+nay09, lob+al08}.  The projected shift in the hydrological cycle will induce even higher temperatures (+3.6\deg C) at the onset of the growing season, already the hottest time of the year, and drier soils when preparations for planting are necessary. 

We can estimate the change in the timing of the Sahel rainy season as we did for PC1: by interpolating from monthly data when rainfall becomes larger than the annual mean (the onset) and when it becomes smaller than the mean (the demise). The mean shifts in a middle-of-the-road scenario (A1B) are, according to this estimate, 5 days in spring and 2 days in the fall; these values are comparable to the 3 day shift estimated for PC1 of the global precipitation, but they indicate a shortening of the rainy season not in the global signal.  Across scenarios (from B1, with the weakest CO2 forcing, to A2, the most aggressive), the spring delay increases with increasing forcing, but the fall delay does not. Although all scenarios show a shortening of the rainy season in the Sahel for the ensemble mean (by 3 days in the B1 and A1B scenarios, 5 in A2), variations across models are large, making the shortening of the rainy season a less robust response to increasing GHG than its delay (see Figure S3).

In the IPCC scenarios a variety of anthropogenic forcings are assumed.  A more
direct assessment of the response to GHG forcing alone is gained by analyzing simulations in which only CO2 is changing and the range of CO2 concentrations is large, strengthening the signal.  In these results, the dependence of the rainy season length on the strength of the GHG forcing appears clearly.  As CO2 concentration grows (increasing to four times the initial values and stabilizing after that), the negative Sahel rainfall anomalies in the first half of the year become increasingly long-lived  (Figure~3a), while positive anomalies do not persist into the dry season. Consequently the rainy season becomes shorter and shorter with increasing GHG, consistent with the scenario simulations.

\begin{figure}
\begin{center}
\includegraphics[]{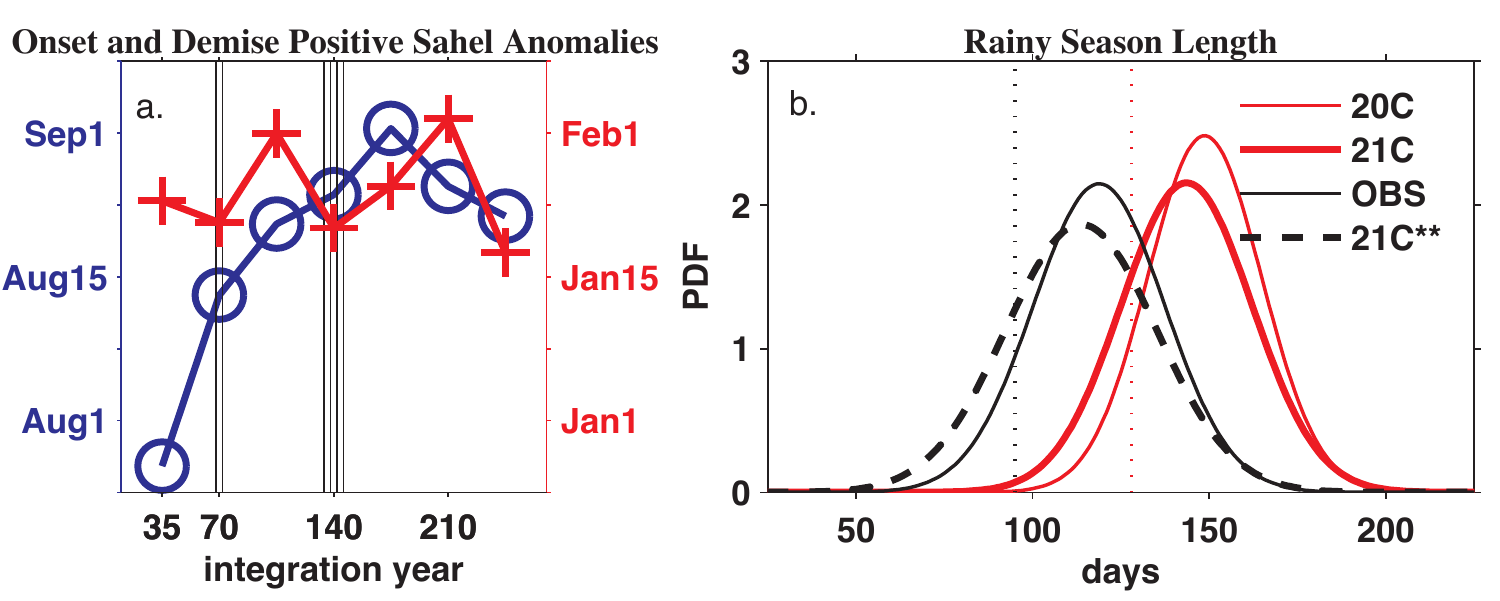}
\caption{{\footnotesize {\bf Change in Sahel Rainy Season Length.} (a.) Dates of zero crossing in Sahel rainfall anomalies in the 1\%to4X simulations, as a function of simulation year. The anomalies are computed as the mean over 20 years around the target dates, minus the long term mean of the control simulation (either pre-industrial or present-day). CO2 doubles at year 70, quadruples at year 140, and remains constant after that. (b.) Probability Density Function (PDF) of Sahel rainy season length in observations (solid black), 20C (thin red), and 21C (thick red). The thick, dashed black line labeled 21C** is a PDF obtained by applying the 21C-20C changes in mean and  standard deviation to the observed PDF, thus it is our best guess of how the observed PDF will change in the 21st century (correcting for model bias). The black (red) dotted vertical lines indicate cumulative probability of 0.1 for OBS (20C). }}
\label{default}
\end{center}
\end{figure}

To begin to quantify what impact a shortening of the rainy season of just a few days would have on the people of the Sahel, we estimate changes in the distribution of the rainy season length from  changes in its mean and variability over the periods 1981-2000 and 2081-2100 in A1B from daily data (Figure~3b). The models overestimate the length of the rainy season (see also Figure~S3), because they increase rainfall between spring and summer too gradually; in the mean, they slightly underestimate the variance, although the observed variance is well within the range spanned by the models. 
The decrease in the mean length of the rainy season (of 5 days, in this dataset) and a slight (insignificant) increase in variance in the 21st century combine to make very short rainy seasons more likely in the future: what once was a one-in-ten year event, is projected to become a one-in-five year event (the change in the mean length alone would make it a one-in-six year event).

The Sahel rainfall changes and the global phase shift of the annual cycle have the same origin.  
A delay in the phase of the annual cycle of surface temperature at high latitudes has been modeled\cite{man+par96} in response to increasing GHG: a consequence of sea ice loss. Where sea ice is present, surface temperature follows local insolation with very little lag, but over open water, the lag increases, due to the higher heat capacity of the oceanic mixed layer. The high latitude oceans show a 21C-20C delay in the rise of surface temperature during spring and summer, but not in early fall, when the oceans are already ice-free at the end of the 20th century. This supports the idea that the high-latitude phase delays are due to sea ice loss.  Recent work suggests in turn that high-latitude forcing can influence tropical climate.  Modeling studies\cite{chi+bit05, kan+al09} and the paleo-record\cite{hau+al01, wan+al08} both indicate that surface temperature changes in the North Atlantic can affect the position of the ITCZs and the monsoon, causing rainfall to migrate further poleward in the warmer hemisphere. By the same mechanisms, a delay in the summertime warming of the northern high latitude would keep rainfall in the SH for longer, and vice versa a delay in the Southern Ocean would keep it to the north.  In any single region, local feedbacks with the land surface might also play a role in making the delay asymmetric, so that the rainy season is shortened. For example, drier soils at the beginning of the rainy season may themselves delay the moistening of the atmosphere necessary to initiate convection; this mechanism has been invoked to explain the lengthening of the dry season in South America\cite{li+al06,set+al09}. 

In the 20th century, observed land surface temperatures show a significant shift towards {\em early} seasons, while the high latitude Atlantic and Pacific show larger, but statistically insignificant, delays\cite{sti+al09}.  The observed advance over mid-latitude continents is not reproduced by the CMIP3 models. This disagreement between models and the observations needs to be explained, but does not by itself invalidate our results for the 21st century.  First, the precipitation delay on which we focus has not emerged in the 20th century, neither in observations, nor in the models (Figure~S4). Second, we expect rainfall to be influenced by SST, especially tropical SST, and little or not at all by mid-latitude land surface temperature: it is likely that phase shifts in the former are governed by different physical process than changes in the latter. 

The similarity of the Sahel response to that projected for South America\cite{li+al06, set+al09}
links both these regional climate changes to a more fundamental, global phase shift in the annual cycle.  The robustness of these results contrasts with the lack of robustness in simulated summertime rainfall changes in the Sahel.  This development strengthens our confidence in climate models, and may result in more useful projections of regional climate changes.  

\bibliography{refs}

\bibliographystyle{naturemag}

{\bf Acknowledgements:} We thank Naomi Naik for her help with the datasets and gratefully acknowledge NOAA support.

\clearpage
\section*{Supplementary Material}

\paragraph{Figure S1: }The evolution of the seasonal cycle of global precipitation is captured by principal component analysis (PCA)\cite{kut67} of the monthly climatology. PCA expands a time-varying field into a set of spatial patterns (empirical orthogonal functions, EOFs) and time series (principal components, PCs) that explain a decreasing fraction of the temporal variance of the original time series.  Figure~S1 shows the second EOF of the global climatological rainfall in the 20C CMIP3 integrations and the 21C-20C changes in PC1. 
\begin{figure}[hbtp]
\begin{center}
\includegraphics[scale=.8]{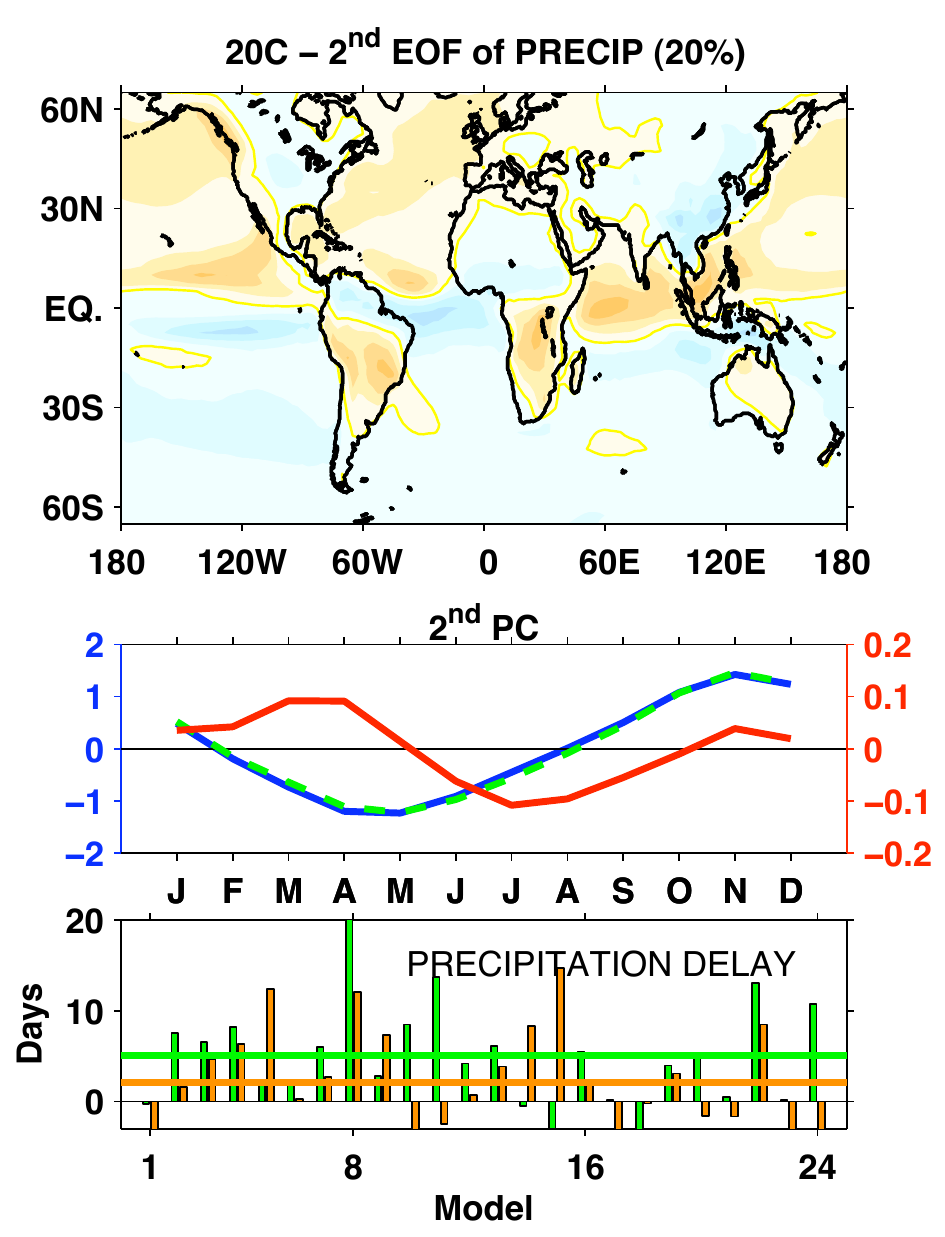}
\caption{{\footnotesize {\bf Supplemental Figure S1.} (Top) Second CMIP3 ensemble mean EOF of the 20th century climatology of precipitation. (Middle)  associated PC2 in the 20C (solid blue) and 21C (dashed green) and 21C-20C difference (red). (Bottom) 21C-20C delay in the timing of the zero crossing in summer (green) and winter (orange) for individual CMIP3 models.}}
\label{EOF2}
\end{center}
\end{figure}

\clearpage

\paragraph{Figure S2: }To verify that the global nature of the delay in the seasonal evolution of rainfall and SST is not an artifact of EOF analysis, we compute the difference in the 21C-20C anomalies  between October and June (Figure S2).  That part of the 21C-20C anomalies common to June and October is removed in this calculation, and only the difference in the response is shown. A delay in the seasonal cycle would then appear in the October minus June 21C-20C anomalies as mimicking the seasonal cycle itself. That is what happens over most of the domain.  Positive SST October-June anomalies are found in the northern hemisphere and negative anomalies in the southern hemisphere, a pattern that resembles the annual cycle (cf. Figure~1), with the exception of the north Indian Ocean and near the Asian coastline. The same resemblance to the annual cycle holds for tropical precipitation anomalies. A  north-south dipole is prominent in the Pacific ITCZ, in the Indian Ocean, in the Americas, and between Northern and Southern Africa. Only in Asia does the pattern become noisy (the expected dipole between Southeast Asia and Indonesia is present,  but anomalies in the Bay of Bengal and over China do not fit the interpretation of a delayed monsoon). We conclude that the delay in the annual cycle is a near global response to GHG forcing. 
\begin{figure}[hbtp]
\begin{center}
\includegraphics[scale=0.75]{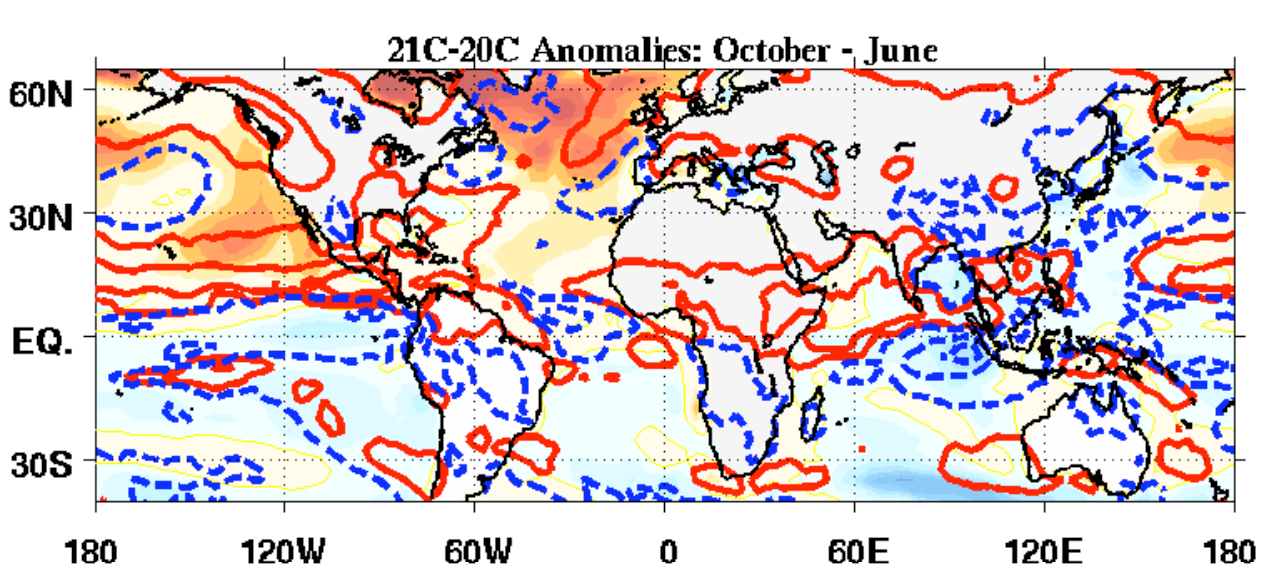}
\caption{{\footnotesize {\bf Supplemental figure S2.} 21C-20C anomalies in October minus 21C-20C anomalies in June. Precipitation is contoured (contours at [-1.5 -.75 -.15 .15 .75  1.5]; dashed blue lines are for negative values, solid red lines are for positive values) and SST is shaded (shading interval is 0.1\deg C, warm colors indicate warm anomalies, the zero line is contoured in yellow). }}
\label{S2}
\end{center}
\end{figure}

\clearpage

\paragraph{Figure S3: } A comparison of observations and CMIP3 model simulations of the timing of the Sahel rainy season (estimated from daily data) is shown in Figure~S3.  The 20C simulations overestimate the length of the rainy season, mostly due to their tendency to anticipate the start date. The amount of variability in start date, end date and length of the rainy season is captured fairly well.
\begin{figure}[hbtp]
\begin{center}
\includegraphics[scale=0.8]{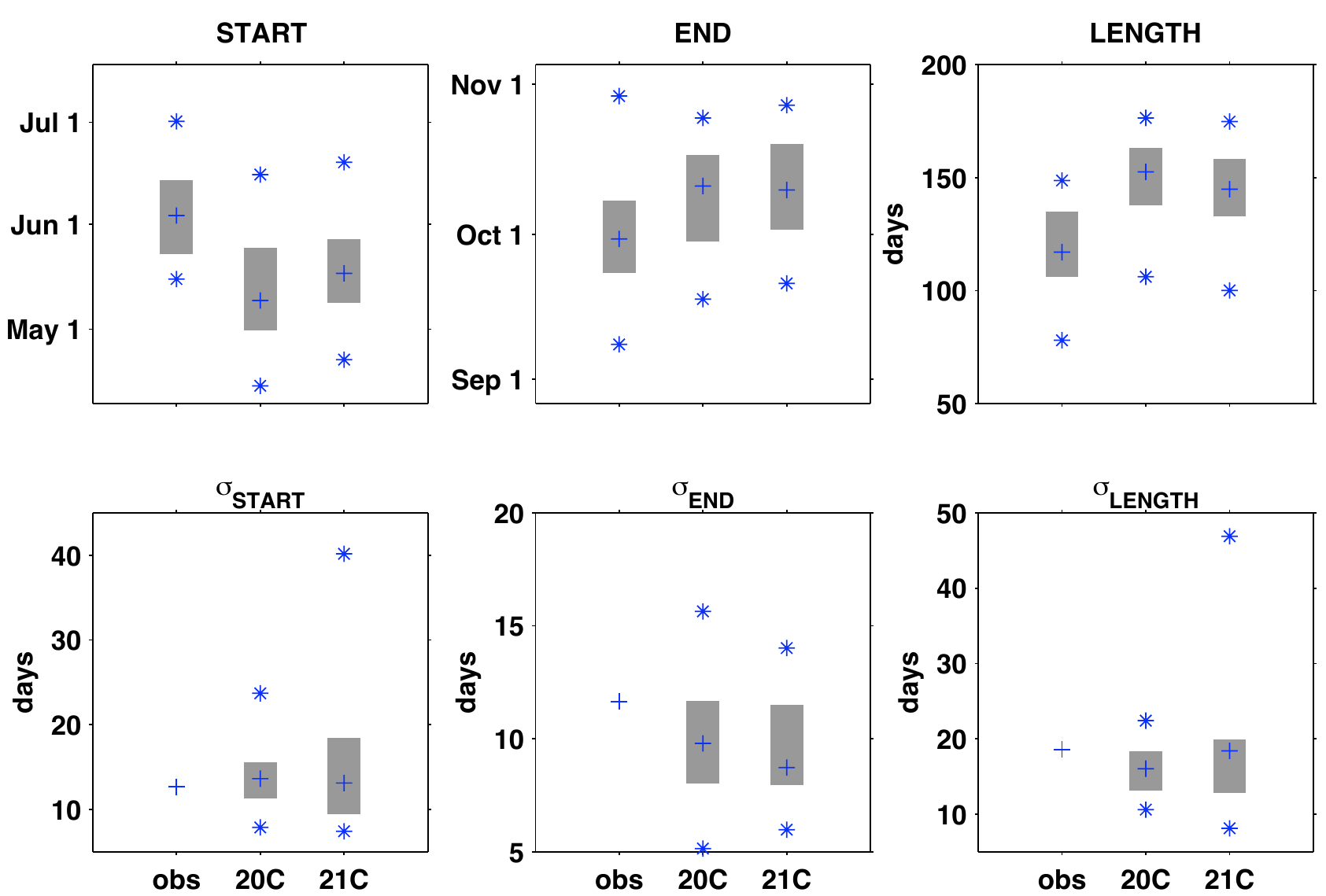}
\caption{{\footnotesize {\bf Supplemental figure S3.} Start date, end date, and length of the rainy season (top) and their standard deviations (bottom)  in observations, 20C and 21C (A1B, daily data)). A plus indicates the median value, the gray bar indicates the 25-to-75 percentiles and asterisks indicate the 2.5 and 97.5 percentiles. For 20C and 21C the statistics are calculated for the model ensemble (each models contributes its mean and standard deviation over 20 year periods). For observations, the standard deviation and percentiles are calculated based on interannual variations for the period 1961-1990.}}
\label{S3}
\end{center}
\end{figure}

\clearpage

\paragraph{Figure S4: }20th century changes in Sahel rainfall show a general decline and no change in seasonality, both in observations and in the CMIP3 models.  This is evident in Figure~S4, which shows  the seasonal evolution of rainfall trends in three observational datasets\cite{GHCN, hulme, xie+arkin} and the 20C minus pre-industrial change in the CMIP3 models.  The observed trend from 1979 has a different seasonal distribution than the longer-term trends, indicating that the response to CO2 projected by the models might be emerging in the observational record as well.
\begin{figure}[hbtp]
\begin{center}
\includegraphics[]{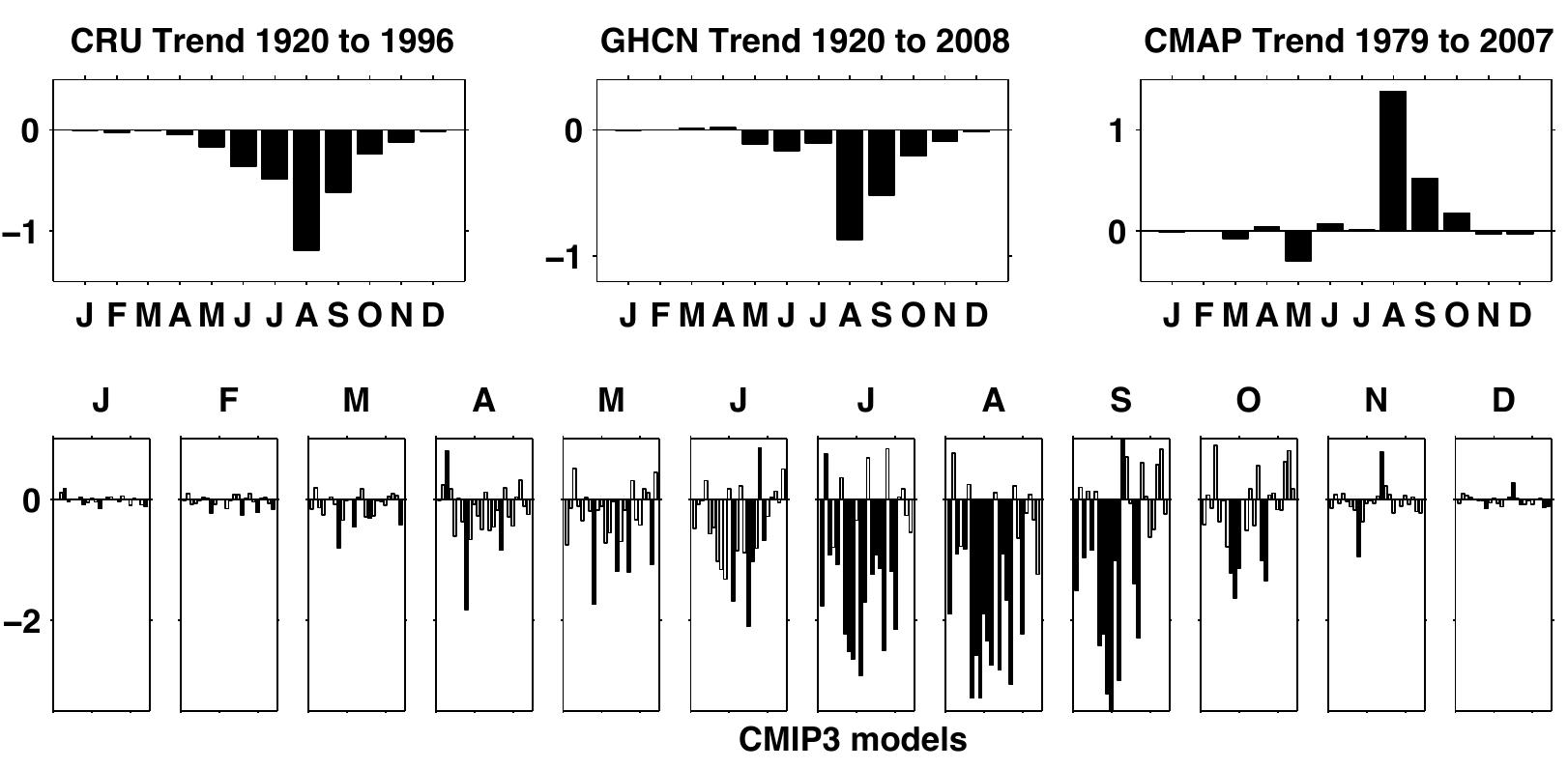}
\caption{{\footnotesize {\bf Supplemental figure S4.} (Top) Observed trend in Sahel rainfall as a function of calender month for 3 different datasets and periods (anomalies are in percent of annual total per decade, where the annual total is calculated from the GHCN dataset and the 1975-1999 period). (Bottom) Sahel rainfall anomalies in the CMIP3 models: difference between the 1975-1999 period in the 20C runs and the long term mean of the control pre-industrial runs. Anomalies that are significant at the 95\% level are solid. Anomalies are in percentage of annual total (the annual total is calculated over the 1975-1999 period from each 20C integration).}}
\label{S4}
\end{center}
\end{figure}

\clearpage



\end{document}